\documentclass[twoside]{article}

\usepackage{qic,epsfig}
\usepackage{graphicx}
\usepackage{subfigure}
\usepackage{cite}
\usepackage{amssymb,amsmath}

\renewcommand{\caption}[1]{\fcaption{#1}}	

\begin{document}
\setlength{\textheight}{8.0truein}    

\runninghead{On the relation between a graph code and a graph state}
            {Yongsoo Hwang and Jun Heo}

\normalsize\textlineskip
\thispagestyle{empty}
\setcounter{page}{1}

\copyrightheading{0}{0}{2003}{000--000}

\vspace*{0.88truein}

\alphfootnote

\fpage{1}

\centerline{\bf
ON THE RELATION BETWEEN}
\centerline{\bf  A GRAPH CODE AND A GRAPH STATE}
\vspace*{0.37truein}
\centerline{\footnotesize
Yongsoo Hwang and Jun Heo\footnote{junheo@korea.ac.kr}}
\vspace*{0.015truein}
\centerline{\footnotesize\it School of Electrical Engineering, Korea University, 
}
\baselineskip=10pt
\centerline{\footnotesize\it Seoul, 136-713, Republic of Korea}
\vspace*{10pt}
\vspace*{0.225truein}
\publisher{(received date)}{(revised date)}

\vspace*{0.21truein}


\abstracts{
A graph state and a graph code respectively are defined based on a mathematical simple graph.
In this work, we examine a relation between a graph state and a graph code both obtained from the same graph, and show that a graph state is a superposition of logical qubits of the related graph code.
By using the relation, we first discuss that a local complementation which has been used for a graph state can be useful for searching locally equivalent stabilizer codes, and second provide a method to find a stabilizer group of a graph code.
}{}{}

\vspace*{10pt}

\vspace*{3pt}
\communicate{to be filled by the Editorial}

\vspace*{1pt}\textlineskip    

\section{Introduction}\label{sec:introduction}

A graph state is a stabilizer state by a stabilizer group, the generators of which are defined by a mathematical graph~\cite{Hein:2004ca,Hein:2005ba,Zeng:2007ja}.
A graph state has been studied in the various field of quantum information science.
First, a graph state plays a prominent role in the design of a codeword stabilized (CWS) quantum code~\cite{Cross:2008jo}.
The error-correcting performance of the code depends on the graph state.
A graph state itself is sometimes referred as an $[[n, 0, d]]$ self-dual graph code~\cite{Danielsen:2008tf}.
Second, a special type of graph states, a cluster state, is a universal resource for an one way quantum computer~\cite{Raussendorf:2001js,Raussendorf:2003ca,Hein:2005ba}.
This quantum computer solves a computational problem by performing continuously single qubit measurements on a cluster state with an appropriately chosen measurement basis. 
Third, one important study about stabilizer states is the local unitary (Clifford) equivalence between stabilizer states.
Since it is well known that an arbitrary stabilizer state can be transformed into a graph state by local Clifford operations, the study can be simplified as the local unitary equivalence between graph states~\cite{VandenNest:2005fp,Zeng:2007ja}.
In addition to these motivations, a graph state can be applicable for secret sharing~\cite{Hein:2005ba,Markham:2008kk}.

In Ref.~\cite{Schlingemann:2001du}, Schlingemann and Werner showed that a quantum error-correcting code can be constructed based on a mathematical graph and the graph-based quantum code called a graph code belongs to stabilizer codes.
In addition, they discussed that the error-correcting ability of a graph code can be easily verified by solving linear equation established by a given graph.
In the follow-up study~\cite{Schlingemann:2001tv}, Schlingemann showed that it is possible to realize a stabilizer code as a graph code. 
How to transform a stabilizer code into a graph code were independently discussed in Refs.~\cite{Grassl:2002ca,VandenNest:2004dj}.
On the other hand, the practicality of a graph code for a fault tolerant quantum computing has been very low because to date a graph code has not almost been described by using the standard stabilizer formalism~\cite{Gottesman:1997ub} based on a Pauli group (or equivalently a symplectic group).

In this work, we investigate a relation between a graph state and a graph code both obtained from the same graph.
In some literature, a graph state is also referred as an $[[n, k=0, d]]$ self-dual graph code~\cite{Danielsen:2008tf}, but in this work we clearly distinguish both.
If logical information can be embedded ($k>0$), then it is a graph code.
Otherwise ($k=0$), it is a graph state.
We believe that a graph state and a graph code defined by the same graph have a close relationship.
The method we employ is a teleportation-like encoding of a graph code~\cite{Beigi:2011fd}, which consists in encoding logical information into a graph state by preparing an initial state that is a tensor product of ancilla qubits and a graph state, applying some Clifford operations, measuring the ancilla qubits and applying additional Clifford operation conditioned on the measurement outcome.
From the investigation, we show that a graph state is a superposition of logical qubits of the related graph code.
By using this relation, we first discuss that a local complementation that is a special Clifford operation acting on a graph state can be useful for searching locally equivalent stabilizer codes. 
Second, we provide how to find a stabilizer group of a graph code.
If a graph code can be written with a stabilizer formalism, its utilization for a fault tolerant quantum computing does not seem awkward anymore.

The rest of this paper is organized as follows.
In section~\ref{sec:preliminary}, we review a graph state and a graph code.
The description about a local complementation is also included in this section.
In section~\ref{sec:relation}, we investigate a relation between a graph state and a graph code both defined by the same graph. 
In section~\ref{sec:local_complementation_graph_code}, we describe how to use a local complementation for graph codes.
In section~\ref{sec:stabilizer_generators}, we discuss how to find a stabilizer group of a graph code.
We finally conclude this paper in section~\ref{sec:conclusion}.

\section{Preliminaries}\label{sec:preliminary}


\subsection{Graph state and local complementation}\label{subset:graph_state}

A \textit{graph state} is a stabilizer state by a stabilizer group, the generators of which are defined by a mathematical graph. 
A vertex and an edge between vertices of a graph correspond to a qubit and a quantum interaction between qubits, respectively.
Given an $n$-vertex graph $G$ (or equivalently an adjacency matrix $\Gamma(G)$), stabilizer generators for the graph state $|G\rangle$ are defined as
\begin{equation}\label{eq:stabilizer_graph_state}
K_j = \sigma_x^{j} \prod_{b\in N_j} \sigma_z^{b},
\end{equation}
where $j=1\sim n$ and $N_j$ is the set of vertices that are adjacent to the vertex $j$.
Note that $\sigma_x^{j}$ is a Pauli-$X$ operator acting on the qubit $j$, and $\sigma_z^{b}$ is similarly defined.
A graph state $|G\rangle$ then is common +1 eigenspace of all these stabilizer generators,
\begin{equation}\label{eq:graph_state}
K_j |G\rangle = |G\rangle,~\textrm{for}~j=1\sim n.
\end{equation}
A graph for a graph state is usually a simple graph that has no self-loops at a vertex and no multiple edges between two vertices. 
In addition, all the edges have the same weight.
A graph with edges of multiple weights can be considered for a non-binary graph state~\cite{Hein:2005ba}.
In this paper, we deal with only a simple graph for a binary graph state (and a binary graph code), and therefore do not use the term ``binary" if there is no confusion.

\begin{figure}[t]
\centerline{
	\subfigure[] {\epsfig{file=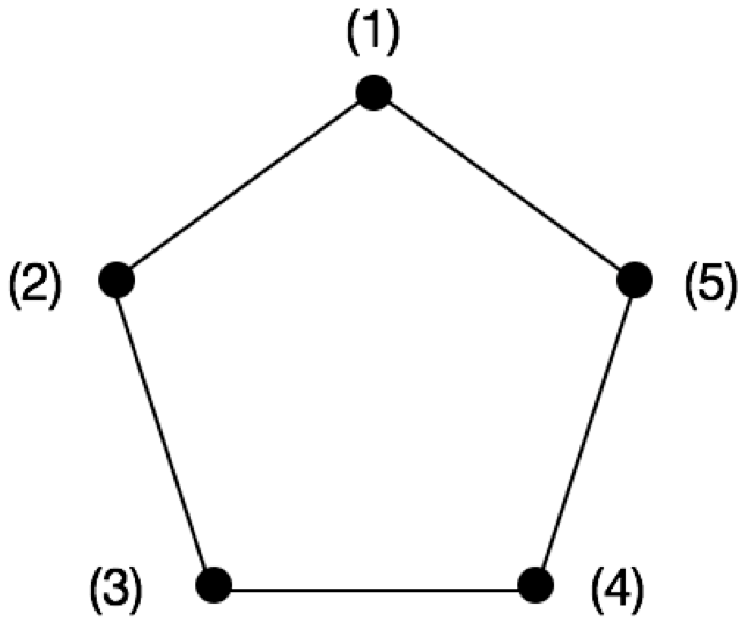,scale=0.4}}
	\label{fig:5_ring_graph_original}
	\subfigure[]{ \epsfig{file=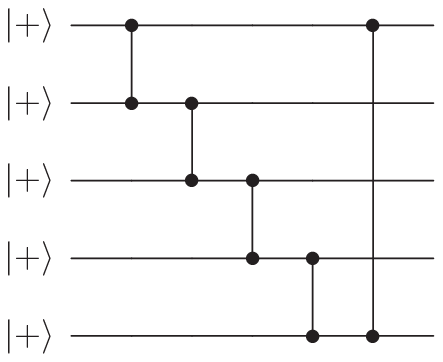,scale=0.7}}
	\label{fig:graph_state_circuit}
}
\fcaption{
(a) A ring graph of length 5, $R_5$. ~(b) A quantum circuit to construct a graph state $|G\rangle$, where $G=R_5$.
}
\label{fig:5_ring_graph}
\end{figure}

A construction of a graph state is very straightforward. 
The application of Controlled-$Z$ ($CZ$) gates to an input state initialized as $|+\rangle^{\otimes n}$ completes the construction~\cite{Hein:2005ba}.
Note that $|+\rangle = \bigl(|0\rangle+|1\rangle\bigr) / \sqrt{2}$. 
The arrangement of $CZ$s is associated with the adjacency matrix $\Gamma(G)$ of a given graph $G$,
\begin{equation}\label{eq:construction_graph_state}
|G\rangle = \Bigl(\prod_{i, j=1}^{n} (CZ^{i,j})^{\Gamma(G)[i,j]}\Bigr) |+\rangle^{\otimes n},
\end{equation}
where $CZ^{i, j}$ is a $CZ$ gate acting on a control qubit $i$ and a target qubit $j$.
Note that $CZ|i\rangle|j\rangle = (-1)^{i\cdot j}|i\rangle|j\rangle$.
Fig.~\ref{fig:5_ring_graph} shows a ring graph of length 5, $R_5$, and a quantum circuit to construct the graph state $|G\rangle$ of $R_5$.
The quantum state at the rightmost of Fig.~\ref{fig:5_ring_graph} (b) is the resulting graph state.

\begin{figure}[t]
\centerline{\epsfig{file=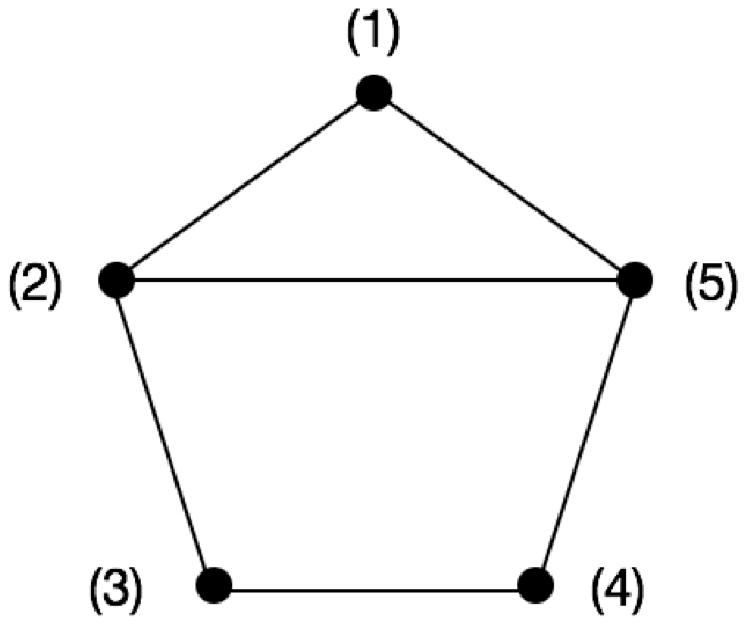,scale=0.4}}
\fcaption
{A graph transformed from $R_5$ by the local complementation at vertex 1, $\tau_1(R_5)$.}
\label{fig:5_ring_graph_lc}
\end{figure}

A \textit{local complementation} is a graph transformation $\tau_v(G)$ such that the subgraph induced by a vertex $v$, $G[N_v]$, is exchanged by its complemented graph $G[N_v]^{c}$ while the other part of the graph remains unchanged~\cite{Bouchet:1993dq}, 
\begin{equation}
\tau_v(G): G[N_v] \mapsto G[N_v]^c.
\end{equation}
Note that $G[N]$ is a subgraph of $G$, which is induced by a vertex set $N$.
The graph transformation can be represented by a matrix computation in terms of the adjacency matrix of a given graph~\cite{VandenNest:2004hn} as
\begin{equation}
\tau_v(G) = \Gamma(G) + \Gamma(G)_v \Gamma(G)_v^T + \Lambda,
\end{equation}
where $\Gamma(G)_v$ is the $v$-th column vector of $\Gamma(G)$ and $\Lambda$ is a diagonal matrix to make $\tau_v(G)[j,j] = 0$ over $j=1\sim n$.
Fig.~\ref{fig:5_ring_graph_lc} shows the transformed graph from $R_5$ of Fig.~\ref{fig:5_ring_graph} (a) by the local complementation at vertex 1.

Surprisingly for the graph operation $\tau_v(G)$ working on a graph $G$, there exists a local Clifford operator $U_v(G)$ acting on the graph state $|G\rangle$ \cite{Hein:2004ca,VandenNest:2004dj,Hein:2005ba},
\begin{equation}\label{eq:local_complementation}
U_v(G) = \sqrt{-i\sigma_x^{(v)}} \prod_{b\in N_v} \sqrt{i\sigma_z^{(b)}},
\end{equation}
where $i$ is the imaginary number, $i^2 = -1$.
Note that $\sqrt{\pm i\sigma_j}$ for $j=x,z,y$ is the $\pi/4$-rotation operator about the $x, z, y$ axis, and its matrix form is represented as 
\begin{equation}
\sqrt{\pm i\sigma_j} = e^{\pm i\frac{\pi}{4} \sigma_j} = \cos \bigl(\frac{\pi}{4}\bigr) I \pm i\sin \bigl(\frac{\pi}{4}\bigr) \sigma_j,
\end{equation}
where 
\begin{equation}
\sigma_x=
\left(
\begin{array}{cc}
0 & 1 \\ 1 & 0
\end{array}
\right),~
\sigma_z=
\left(
\begin{array}{cc}
1 & 0 \\ 0 & -1
\end{array}
\right),~
\sigma_y=
\left(
\begin{array}{cc}
0 & -i \\ i & 0
\end{array}
\right), 
\end{equation}
and $I$ is the identity matrix of size $2\times 2$.
As an example, the local Clifford operator for $\tau_1(R_5)$ is expressed as
\begin{equation}
U_1(R_5) = \sqrt{-i\sigma_x^{(1)}} \sqrt{i\sigma_z^{(2)}} \sqrt{i\sigma_z^{(5)}}.
\end{equation}

By a local complementation $U_v(G)$, a graph state $|G\rangle$ is transformed into a locally Clifford equivalent graph state $|G'\rangle$, 
\begin{equation}
|G'\rangle = U_v(G)|G\rangle,
\end{equation}
which is by definition a graph state by the graph $G'$ that is transformed from $G$ by $\tau_v(G)$.
As a consequence, it is believed that by applying successive local complementations to a graph state $|G\rangle$, one can find a complete set of local Clifford equivalent graph states of $|G\rangle$~\cite{VandenNest:2004dj}.


\subsection{Graph code}\label{subsec:graph_code}

In Ref.~\cite{Schlingemann:2001du}, a graph code is constructed by a graph and a finite abelian group.
The authors described a graph code as an isometry from input information qubits to output physical qubits, and the isometry is defined as an integral over both qubits. 
Each qubit is associated with a vertex of a graph.
In this work, we review a graph code in terms of a graph structure only because we focus on a binary graph code based on the abelian group $\mathbb{F}_2$.

At the beginning of this work, we mentioned that we would investigate a relation between a graph code and a graph state both derived from the same graph.
However, the mention is half correct and half not, because a graph for a graph code is not exactly same as that for a graph state. 
As mentioned above, a graph code is defined by an isometry from information qubits to physical qubits, and therefore a graph for a graph code has to represent the isometry by itself.
Which means that a graph for a graph code consists of two distinct vertex sets, $V_{in}$ and $V_{out}$, of the orders $k$ and $n$.
To conclude, an $(n+k)$-vertex graph is required to design a graph code that corresponds to an $[[n, k, d]]$ stabilizer code while an $n$-vertex graph defines a graph state.
Note that even though vertices are classified into two sets, a connection between two vertices in the set $V_{out}$ is allowed, namely, the graph is not bipartite.

What we mentioned at the beginning therefore means a relation between a graph code and a graph state where the graph state is defined by the subgraph induced by $V_{out}$ of a graph for the graph code.
Throughout this work, we indicate that a graph for a graph state is the induced subgraph by $V_{out}$ of the graph for a graph code.
To avoid any confusion in graphs, we call an $n$-vertex graph a \textit{normal} graph denoted by $G$, and an $(n+k)$-vertex graph an \textit{extended} graph denoted by $G^{Ext}$.
Note that in Ref.~\cite{Cafaro:2014vm}, the authors use the term ``extended graph" to indicate a graph for a graph code, which is exactly the same as our extended graph.

An extended graph for a graph code can be given in the beginning~\cite{Schlingemann:2001du}, or can be found from a given normal graph~\cite{Grassl:2002ca}.
How to derive a graph state (therefore a graph) from a stabilizer code, described independently in Refs.~\cite{Grassl:2002ca,VandenNest:2004dj}, are almost the same: applying \textit{local Clifford operations} and an \textit{invertible linear operation} to the stabilizer code. 
Note that prior to the application of these operations, a $[[n, k, d]]$ stabilizer code has to be transformed to a related $[[n, 0, d]]$ stabilizer code by measuring additional $k$ commuting operators\footnote{These additional $k$ operators have to commute with the existing $n-k$ stabilizer generators. The usual choice for them are $k$ logical $Z$ operators.}.

\begin{figure}[t]
\centerline{\epsfig{file=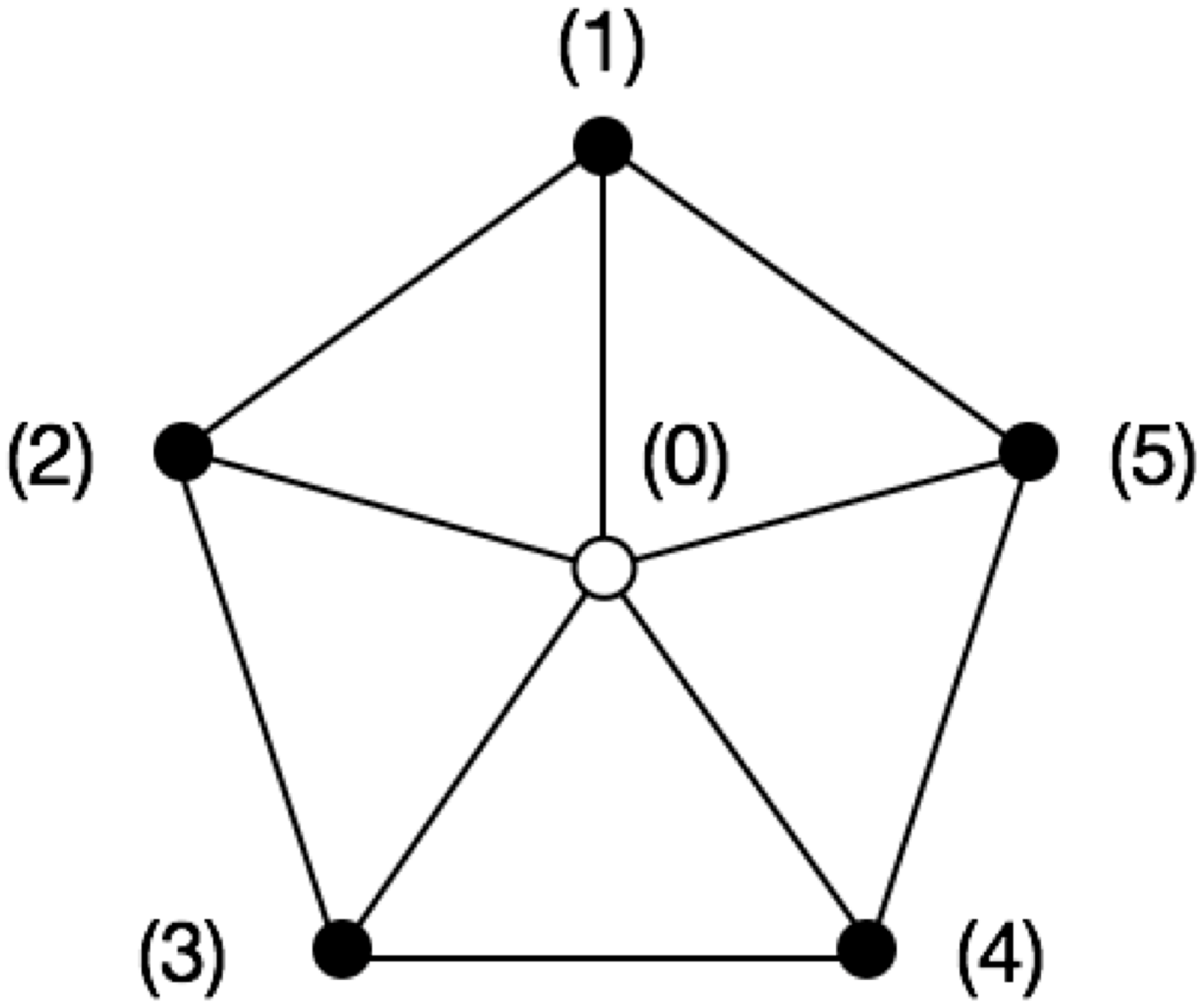,scale=0.25}}
\fcaption{
	The extended graph of the ring graph $R_5$ with one input vertex (unfilled vertex), $k=1$.
	Since the input vertex is connected to all the other vertices, a logical information in the input vertex is spread out over all the output qubits. 
}
\label{fig:5_ring_graph_extended}
\end{figure}

Given a normal graph $G$, the adjacency matrix of the extended graph $G^{Ext}$ is described as
\begin{equation}
\Gamma(G^{Ext}) = \left(
\begin{array}{cc}
0 & B \\
B^{T} & \Gamma(G)
\end{array}
\right),
\end{equation}
where $\Gamma(G)$ is an adjacency matrix of $G$.
The matrix $B$ of size $k\times n$ shows connections between $k$ input vertices (qubits) and $n$ output vertices (qubits), namely if $B[i,j]=1$, then the vertex $v^{in}_i$ is connected to the vertex $v^{out}_j$, where $v^{in}_i\in V_{in}$ and $v^{out}_j\in V_{out}$.
As indicated in Refs.~\cite{Grassl:2002ca,Cafaro:2014vm}, $B$ and $\Gamma(G)$ are orthogonal, $B\cdot \Gamma(G)=0$.
Fig.~\ref{fig:5_ring_graph_extended} shows an extended graph from $R_5$ whose $B$ matrix is $(1~1~1~1~1)$.
This graph defines a $[[5, 1]]$ graph code that performs an encoding by spreading 1-qubit logical information in the input vertex $0$ over all the output vertices.

\section{A relation between a graph state and a graph code}\label{sec:relation}

\begin{figure}[t]
\centerline{\epsfig{file=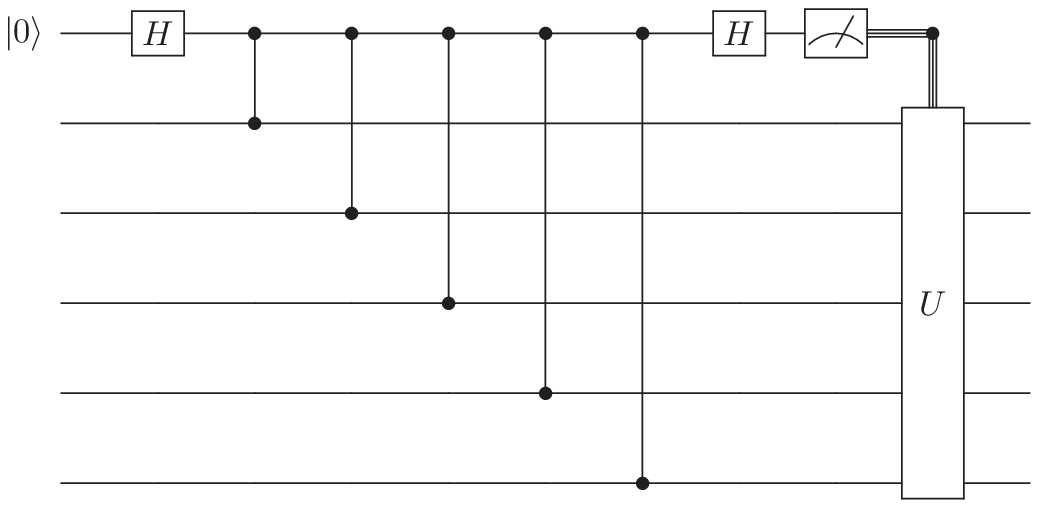,scale=0.7}}
\fcaption{
The encoder of a graph code associated with $R_5$.
The input state of the lower 5 qubits at the leftmost is the graph state of $R_5$.}
\label{fig:graph_code_encoder}
\end{figure}

In this section, we investigate a relation between a graph state and a graph code.
To this end, we first show how to encode a logical information with a graph code, which is achieved by a teleportation-like method that consists of preparing an initial state, applying $CZ$ operations, performing measurements, and applying additional Clifford operation conditioned on the measurement outcome~\cite{Beigi:2011fd}.
The initial state is a tensor product of $k$ information qubits and a graph state we constructed with a given normal graph $G$.
\begin{equation}
|\psi\rangle_{init} = |c_1\cdots c_k\rangle \otimes |G\rangle,
\end{equation}
where $|c_1\cdots c_k\rangle = X_{1}^{c_1}\cdots X_k^{c_k}|0\cdots 0\rangle$.
Note that $X_i$ is a Pauli $X$ operator acting on the $i$-th qubit.
After the applications of $H$ gates on the ancilla qubits, $CZ$ gates according to the matrix $B$, and again $H$ gates on the ancilla qubits in series, one measures the ancilla qubits.
\begin{equation}
|\psi\rangle_{final} = \Bigl(H^{\otimes k}\otimes I^{\otimes n}\Bigr)\cdot\widetilde{CZ}\cdot \Bigl(H^{\otimes k}\otimes I^{\otimes n}\Bigr) |\psi\rangle_{init},
\end{equation}
where $\widetilde{CZ}$ is
\begin{equation}
\widetilde{CZ} = \prod_{i, j} (CZ^{i, j})^{B[i,j]},
\end{equation}
over $1\leq i\leq k$ and $1\leq j\leq n$, and $|\psi\rangle_{final} = |m_1\cdots m_k\rangle \otimes |\phi\rangle$.
Since the matrix $B$ shows connections between input vertices and output vertices, the application of $\widetilde{CZ}$ introduces quantum interactions between input qubits and output qubits.
If the measurement outcome $|m_1\cdots m_k\rangle$ is $|0\cdots 0\rangle$, then the quantum state $|\phi\rangle$ corresponds to a logical qubit $|c_1\cdots c_k\rangle_L$.
Otherwise, one has to apply $\bar{X}_1^{m_1}\cdots \bar{X}_k^{m_k}$ to $|\phi\rangle$ conditioned on the measurement outcome $|m_1\cdots m_k\rangle$, where $\bar{X}_j$ is a logical $X$ operator of the related graph code.
As an example, Fig.~\ref{fig:graph_code_encoder} shows an encoding circuit of the graph code we derived from $R_5$.
The matrix $B$ in this case is $B=(1~1~1~1~1)$ as mentioned before.

Let us examine each step of the encoding process we have shown with an assumption $k=1$.
Here we suppose that the information to encode is $|0\rangle$, and a graph state $|G\rangle$ is arbitrary.
The initial quantum state then is $|0\rangle\otimes |G\rangle$.
After the applications of $H$ and $CZ$, the quantum state will be
\begin{equation}
\frac{1}{\sqrt{2}} \Bigl(|0\rangle|G\rangle + \widetilde{CZ}|1\rangle|G\rangle \Bigr).
\end{equation}
Since $CZ^{1, j}|1\rangle|G\rangle$ equals to $|1\rangle Z^{j}|G\rangle$ for $j>1$, the above state can be written as
\begin{equation}
\frac{1}{\sqrt{2}} \Bigl(|0\rangle|G\rangle + |1\rangle \widetilde{Z}|G\rangle \Bigr),
\end{equation}
where 
\begin{equation}\label{eq:logical_z_graph_code}
\widetilde{Z} = \prod_{j=1}^{n} \bigl( Z^{(j)} \bigr)^{B[j]}.
\end{equation}
After the application of $H$ again, the state will be 
\begin{equation}
|0\rangle \Bigl( \frac{I+\widetilde{Z}}{2} \Bigr)|G\rangle + |1\rangle \Bigl(\frac{I-\widetilde{Z}}{2} \Bigr)|G\rangle.
\end{equation}
The logical zero qubit $|0\rangle_L$ therefore is
\begin{equation}\label{eq:logical_zero}
|0\rangle_L = \Bigl( \frac{I+\widetilde{Z}}{2}\Bigr)|G\rangle.
\end{equation}
Since the operator $(1+\widetilde{Z})/2$ is the projector onto the $+1$ eigenspace of $\widetilde{Z}$, one can say that $|0\rangle_L$ is a superposition of the observables of $|G\rangle$, which belong to the $+1$ eigenspace of $\widetilde{Z}$.
Since by definition a Hermitian Pauli operator that squares to $+I$ has eigenvalues $+1$ and $-1$ only, the other observables that will be projected onto the $-1$ eigenspace of $\widetilde{Z}$ compose $|1\rangle_L$.
As a consequence a graph state $|G\rangle$ is a superposition of the logical qubits $|0\rangle_L$ and $|1\rangle_L$ of the graph code,
\begin{equation}\label{eq:graph_state_superposition}
|G\rangle = |0\rangle_L + |1\rangle_L.
\end{equation}

Let us denote a logical $Z$ operator of a graph code by $\bar{Z}$.
Then, 
\begin{equation}\label{eq:graph_state_logical_z}
\bar{Z} |G\rangle = |0\rangle_L - |1\rangle_L.
\end{equation}
From Eqs~(\ref{eq:graph_state_superposition}) and (\ref{eq:graph_state_logical_z}),
\begin{equation}
\Bigl( \frac{1+\bar{Z}}{2} \Bigr) |G\rangle = |0\rangle_L.
\end{equation}
By a comparison with Eq.~(\ref{eq:logical_zero}), we can say that $\bar{Z}$ is equivalent with $\widetilde{Z}$, namely the operator $\widetilde{Z}$ that derived from the matrix $B$ corresponds to the logical $Z$ operator $\bar{Z}$ of a graph code.
To conclude, when $k=1$, a graph state is a superposition of logical qubits of the relevant graph code, and the logical $Z$ operator $\bar{Z}$ is defined by the matrix $B$.

What happens when $k>1$ ?
Without loss of generality, we show the case of $k=2$.
From the matrix $B$ of size $2\times n$, one obtains the following operators
\begin{equation}
\widetilde{Z}_i = \prod_{j=1}^{n} \Bigl( Z^{(j)}\Bigr)^{B[i,j]} ,
\end{equation}
for $i=1, 2$ and $j=1\sim n$.
Suppose that we encode a logical information $|00\rangle$.
The initial state then is a tensor product of $|00\rangle$ and an arbitrary graph state $|G\rangle$.
By the applications of $H$ gates on the ancilla qubits, $\widetilde{Z}_i$ according to the ancilla qubits $|c_i\rangle$, and again $H$ gates on the ancilla qubits, the initial state is transformed into the following state
\begin{equation}
\sum_{c_1, c_2}|c_1,c_2\rangle \Bigl(\frac{I+(-1)^{c_1}\widetilde{Z}_1}{2}\Bigr)\Bigl(\frac{I+(-1)^{c_2}\widetilde{Z}_2}{2}\Bigr) |G\rangle,
\end{equation}
where $c_1,c_2\in\{0,1\}$.
If the measurement outcome, performed on the ancilla qubits, is $|00\rangle$, then the quantum state 
\begin{equation}
\Bigl( \frac{I+\widetilde{Z}_1}{2}\Bigr)\Bigl( \frac{I+\widetilde{Z}_2}{2}\Bigr)|G\rangle
\end{equation}
corresponds to the logical qubit $|00\rangle_L$.
The logical qubit $|00\rangle_L$ is a $+1$ common eigenspace of all $\widetilde{Z}_i$s from the observables of $|G\rangle$.
If the measurement outcome is $|m_1,m_2\rangle$, then controlled logical $X$ operator $\prod \bar{X}_i^{m_i}$ should be applied to find $|00\rangle_L$.
This means that one can find $|c_1,c_2\rangle_L$ by projecting the graph state $|G\rangle$ with a projector 
\begin{equation}
\prod_{i=1}^{2} \Bigl( \frac{I+(-1)^{c_i}\widetilde{Z}_i}{2}\Bigr).
\end{equation}
even though one prepared the initial state $|00\rangle\otimes|G\rangle$.

From the cases $k=1, 2$, we can say that a quantum state $|0\cdots 0\rangle|G\rangle$ is transformed into
\begin{equation}
\sum_{c_1, \cdots, c_k} |c_1\cdots c_k\rangle \prod_{i=1}^{k}\Bigl( \frac{I+(-1)^{c_i}\widetilde{Z}_k}{2}\Bigr) |G\rangle,
\end{equation}
by the applications of $H$s, $\widetilde{Z}$s, and again $H$s.
After performing measurements on the ancilla qubits, if the measurement outcome is $|c_1\cdots c_k\rangle$, then the quantum state of the unmeasured qubits corresponds to $|c_1\cdots c_k\rangle_L$.
If necessary, applying controlled logical $X$ operators conditioned on the measurement outcome completes finding a certain logical qubit.

By applying the argument made in the case $k=1$, one can say that $\widetilde{Z}_i$ is the $i$-th logical $Z$ operator $\bar{Z}_i$ of a graph code.
Therefore, in the remainder of this work, we will denote it by $\bar{Z}_i$ if there is no confusion.
To conclude, a graph state $|G\rangle$ is a superposition of logical qubits of the related graph code,
\begin{equation}
|G\rangle = \sum_{c_1,\cdots, c_k} |c_1\cdots c_k\rangle_L.
\end{equation}
When a logical information $|0\cdots 0\rangle$ is given in the beginning, one can find the logical qubit $|c_1\cdots c_k\rangle_L$ from the measurement outcome $|c_1\cdots c_k\rangle$ after the applications of operations.
\begin{equation}
|c_1\cdots c_k\rangle_L = \prod_{j=1}^{k} \Bigl( \frac{I+(-1)^{c_j}\bar{Z}_j}{2} \Bigr)|G\rangle.
\end{equation}

\section{Local complementation for a graph code}\label{sec:local_complementation_graph_code}

In the previous section, we showed that a graph state is a superposition of logical qubits of the associated graph code, $|G\rangle = \sum |c_1 \cdots c_k\rangle_L$.
Therefore, a local complementation proposed for a graph state can be applied to a graph code.
To be exact, the transformation operation is applied to a graph state, and then the transformed graph state is projected to a logical qubit of the related graph code.

Let us assume that two graph states $|G_1\rangle$ and $|G_2\rangle$ are local Clifford equivalent under a local complementation transformation, $U(G_1)_v|G_1\rangle = |G_2\rangle$.
Let us denote the associated graph code of the graph state $|G_i\rangle$ by $\mathcal{C}_{G_i}$.
From the relation
\begin{equation}
U_v(G_1) \Bigl( \sum |c_1\cdots c_k\rangle_L^{\mathcal{C}_1} \Bigr) = \sum |c_1 \cdots c_k\rangle_L^{\mathcal{C}_2} 
\end{equation}
one can read that the logical qubits of $\mathcal{C}_{G_1}$ and $\mathcal{C}_{G_2}$ are local Clifford equivalent because $U_v(G_1)$ is composed of several single qubit Pauli operators (See Eq.~(\ref{eq:local_complementation})).
Note that $|c_1\cdots c_k\rangle^{\mathcal{C}_i}_L$ is a logical qubit of a graph code $\mathcal{C}_{G_i}$.

The relation between the logical qubits, $|c_1\cdots c_k\rangle^{\mathcal{C}_1}_L$ and $|c_1\cdots c_k\rangle^{\mathcal{C}_2}_L$, should be described through a local complementation transformation between graph states as
\begin{equation}\label{eq:logical_qubits_conversion}
|c_1\cdots c_k\rangle^{\mathcal{C}_1}_L \rightarrow |G_1\rangle \rightarrow |G_2\rangle \rightarrow |c_1\cdots c_k\rangle^{\mathcal{C}_2}_L.
\end{equation}
As an example, we show the transformation from $|0\cdots 0\rangle_L^{\mathcal{C}_1}$ to $|0\cdots 0\rangle_L^{\mathcal{C}_2}$.
The first step $|0\cdots 0\rangle^{\mathcal{C}_1}_L \rightarrow |G_1\rangle$ can be done via the applications of $H$ gate and controlled logical $X$ gates of the graph code $\mathcal{C}_{G_1}$. 
The initial state is a tensor product of ancilla qubits $|+\rangle^{\otimes k}$ and the logical qubit $|0\cdots 0\rangle_L^{\mathcal{C}_1}$,
\begin{eqnarray}
|\phi\rangle_{init} &=& |+\rangle^{\otimes k} |0\cdots 0\rangle_L^{\mathcal{C}_1} \\
&=&\sum_{c_i\in\{0, 1\}} |c_1\cdots c_k\rangle |0\cdots 0\rangle_L^{\mathcal{C}_1}.
\end{eqnarray}
After applications of controlled-$\bar{X}_i$ conditioned on the ancilla qubit $c_i$, the initial state is transformed into 
\begin{eqnarray}
|\phi\rangle_{mid} &=& \bar{X}^{c_1\cdots c_k} |\phi\rangle_{init}\\
 &=& \sum |c_1\cdots c_k\rangle  |c_1\cdots c_k\rangle_L^{\mathcal{C}_1}.
\end{eqnarray}
After the application of $H$ gates to the ancilla qubits again, one measures the ancilla qubits.
\begin{eqnarray}
|\phi\rangle_{final} &=& \Bigl(H^{\otimes k}\otimes I^{\otimes n}\Bigr) |\phi\rangle_{mid} \\
&=& \sum |m_1\cdots m_k\rangle \prod_{j=1}^{k}\bar{Z}_j^{m_j} |G_1\rangle, 
\end{eqnarray}
where $|G_1\rangle = \sum  |c_1\cdots c_k\rangle_L^{\mathcal{C}_1}$.
If the measurement outcome is $|0\cdots 0\rangle$, then the unmeasured state will be the graph state $|G_1\rangle$.
Otherwise, the controlled logical $Z$ operator $\bar{Z}_j$ has to be applied according to the measurement outcome $m_j$.

Thereafter, $|G_1\rangle$ is transformed into $|G_2\rangle$ by the local complementation $U_v(G_1)$, and $|G_2\rangle$ is projected to $|0\cdots 0\rangle_L^{\mathcal{C}_2}$ of the graph code $\mathcal{C}_{G_2} $by the projection as
\begin{equation}\label{eq:projection_g2}
|0\cdots 0\rangle_L^{\mathcal{C}_2} = \prod_{j=1}^{k} \Bigl( \frac{I+\bar{Z}_j}{2}\Bigr) |G_2\rangle,
\end{equation}
where $|G_2\rangle = U_v(G_1)|G_1\rangle$, and 
\begin{equation}\label{eq:back_to_g1}
|G_1\rangle = (\bar{Z}^{c_1\cdots c_k}) (H^{\otimes k}I^{\otimes n} ) (\bar{X}^{c_1\cdots c_k}) (|+\rangle^{\otimes k}|0\cdots0\rangle_L^{\mathcal{C}_1}).
\end{equation}
Note that $\bar{Z}_j$ used in Eq.~(\ref{eq:projection_g2}) is the logical $Z$ operator for the graph code $\mathcal{C}_{G_2}$, and the other logical $X$ or $Z$ operators mentioned in this section are for $\mathcal{C}_{G_1}$.
The controlled logical operations $\bar{X}^{c_1\cdots c_k}$ is a product of logical $X$ operators $\bar{X}_j$ conditioned on the measurement outcome $c_j$
\begin{equation}
\bar{X}^{c_1\cdots c_k} = \prod \bar{X}_j^{c_j},
\end{equation}
and $\bar{Z}^{c_1\cdots c_k}$ is similarly defined, and the last controlled logical $Z$ operators of Eq.~(\ref{eq:back_to_g1}) has to be performed according to the measurement outcome.

On the surface, the readers may think that $|c_1\cdots c_k\rangle_L^{\mathcal{C}_2}$ can be derived by applying a local complementation transformation $U_v(G_1)$ directly to $|c_1\cdots c_k\rangle_L^{\mathcal{C}_1}$. 
Unfortunately, this is not true. 						
Let us suppose that $k=1$.
For $U_v(G_1)|0\rangle_L^{\mathcal{C}_1} = |0\rangle_L^{\mathcal{C}_2}$, the following equation has to hold true,
\begin{equation}
\Bigl(\frac{I+\bar{Z}}{2}\Bigr) U_v(G_1) (|0\rangle_L^{\mathcal{C}_1} + |1\rangle_L^{\mathcal{C}_1}) = U_v(G_1)|0\rangle_L^{\mathcal{C}_1},
\end{equation}
where 
\begin{equation}
|0\rangle_L^{\mathcal{C}_1} = \Bigl(\frac{I+\bar{Z}}{2}\Bigr) (|0\rangle_L^{\mathcal{C}_1} + |1\rangle_L^{\mathcal{C}_1}). 
\end{equation}
But, it is not because $\bar{Z}$ and $U_v(G_1)$ do not mutually commute in general.
To conclude, a local complementation transformation between logical qubits has to be done by a local complementation between the graph states $|G_1\rangle$ and $|G_2\rangle$ as described in Eq.~(\ref{eq:logical_qubits_conversion}).

\section{How to find stabilizer group of a graph code}\label{sec:stabilizer_generators}

As indicated in Refs.~\cite{Schlingemann:2001du,Schlingemann:2001tv}, a graph code has an advantage of simplification in checking the error-correcting capability for a certain quantum error, however its utilization for a fault tolerant quantum computing seems awkward because to date a graph code has not almost been described by using the standard stabilizer formalism.
For a variety of reasons, we believe that it is necessary to represent a graph code with the stabilizer formalism~\cite{Gottesman:1997ub}. 
In this section, we discuss about how to find stabilizer generators of a graph code.

For the sake of easy understanding, we first discuss the case $k=1$.
Given a normal graph $G$, one can easily find a stabilizer group $\mathcal{S}_{|G\rangle}$ of the graph state $|G\rangle$, which is generated by the stabilizer generators $K_j$ over $j=1\sim n$ described in Eq.~(\ref{eq:stabilizer_graph_state}).
Since an extended graph for a graph code can be derived from the normal graph, the logical $Z$ operator of the  graph code $\mathcal{C}_{G}$, which is defined by the matrix $B$, also can be found by following Eq.~(\ref{eq:logical_z_graph_code})

Suppose that a stabilizer group $\mathcal{S}_\mathcal{C}$ of a graph code $\mathcal{C}_G$ is generated by a set of $r$ stabilizer generators, $\mathcal{S}_\mathcal{C} = \langle g_1, \cdots, g_r \rangle$ where $r=n-k$.
One can find $r$ stabilizer generators by using the following three facts.
First, a stabilizer of the graph code $\mathcal{C}_G$ also stabilizes the graph state $|G\rangle$,
\begin{equation}
U|G\rangle = U|0\rangle_L + U|1\rangle_L = |0\rangle_L + |1\rangle_L = |G\rangle,
\end{equation}
where $U\in \mathcal{S}_\mathcal{C}$.
Second, the logical $Z$ operator $\bar{Z}$ of $\mathcal{C}_G$ does not belong to the stabilizer group $\mathcal{S}_{|G\rangle}$ of the graph state $|G\rangle$, that is, there exist several graph state stabilizers $K_j$s such that $\{\bar{Z}, K_j\}=0$. 
Third, $\bar{Z}$ that belongs to a normalizer of $\mathcal{C}_G$ commutes with a stabilizer generator of $\mathcal{C}_G$, $[\bar{Z}, g_j]=0$.

Suppose that a weight of $\bar{Z}$ is $w$, where a weight of an $n$-qubit Pauli operator is defined as the number of non-identity single qubit Pauli operators involved in the operator.
To find stabilizer generators of $\mathcal{C}_G$, one first needs to divide the stabilizer generators $K=\{K_1, \cdots, K_n\}$ of the graph state $|G\rangle$ into two sets $K^1$ and $K^2$ defined as 
\begin{equation}
\begin{array}{l}
K^1 = \{M | [M, \bar{Z}] = 0,~\textrm{where}~M\in K\} \\
K^2 = \{N | \{N, \bar{Z}\} = 0,~\textrm{where}~N\in K\}
\end{array}.
\end{equation}
From the definitions of $K_j$ and $\bar{Z}$, the orders of both sets are respectively $|K^1|=n-w$ and $|K^2|=w$.
By the above-mentioned third fact, all elements of the set $K^1$ belong to $\mathcal{S}_\mathcal{C}$.
In addition, $K_{j}$ is linearly independent, the elements of $K^1$ can become stabilizer generators.
One thus has found $n-w$ stabilizer generators.

The remaining $w-k$ generators can be obtained  by exploiting the following the relation, $n$-qubit Pauli operators $A$ and $BC$ mutually commute even when $\{A, B\} = \{A, C\} = 0$, 
\begin{equation}\label{eq:commuting_with_anticommuting}
A BC = -BAC = BC A. 
\end{equation}
Therefore, some multiplication products of even elements in $K^2$ are commuting with $\bar{Z}$.
Among them, finding $w-k$ linearly independent elements completes the stabilizer group $\mathcal{S}_\mathcal{C}$ of the graph code $\mathcal{C}_G$.
When $k=1$, this process is very straightforward: pick a pivot element $N_p$ randomly and make a multiplication product of $N_p$ and $N_j$, $N_p\cdot N_j$ for $j=1\sim w$ and $j\neq p$.

As an example, let us show how to find stabilizer generators of the graph code associated with the graph $R_5$.
The stabilizer generators of the graph state are as follows,
\begin{equation}
K = \left(
\begin{array}{ccccc}
X & Z & I & I & Z \\
Z & X & Z & I & I \\
I & Z & X & Z & I\\
I & I & Z & X & Z\\
Z & I & I & Z & X
\end{array}
\right).
\end{equation}
As shown in section~\ref{subsec:graph_code}, for $R_5$, $B=(1~1~1~1~1)$, and therefore the set $K^1$ is an empty set because $\bar{Z}=ZZZZZ$.
Then, one has to find all stabilizer generators from multiplication products of even elements from $K^2$. 
By picking $K_1=XZIIZ$ as a pivot, we can find the stabilizer generators by making a product $K_1\cdot K_j$ for $j=2\sim 5$ as
\begin{equation}
\mathcal{G}=\left(
\begin{array}{ccccc}
Y & Y & Z & I & Z \\
X & I & X & Z & Z \\
X & Z & Z & X & I \\
Y & Z & I & Z & Y
\end{array}
\right).
\end{equation}
Even though we have not explicitly mentioned in this paper, the graph $R_5$ of Fig.~\ref{fig:5_ring_graph} (a) is a graph realization of the well-known $[[5,1,3]]$ stabilizer code~\cite{Schlingemann:2001du} defined by the following stabilizer generators
\begin{equation}
\mathcal{G}^{[[5,1,3]]} = 
\left(
\begin{array}{ccccc}
X & Z & Z & X & I \\
I & X & Z & Z & X \\
X & I & X & Z & Z \\
Z & X & I & X & Z
\end{array}
\right).
\end{equation}
The readers easily can see that these standard stabilizer generators are equivalent with those we found above.

In the original work of a graph code~\cite{Schlingemann:2001du,Schlingemann:2001tv}, how to correct errors with a graph code was not explicitly described.
From now on, one can perform a syndrome measurement for a passive quantum error correction. 
As an example, in case of the graph code by $R_5$, when the error $``IIXII"$ is occurred, one obtains the error syndrome $(-1, +1, -1, +1)$.

\begin{figure}[t]
\centering{
	\epsfig{file=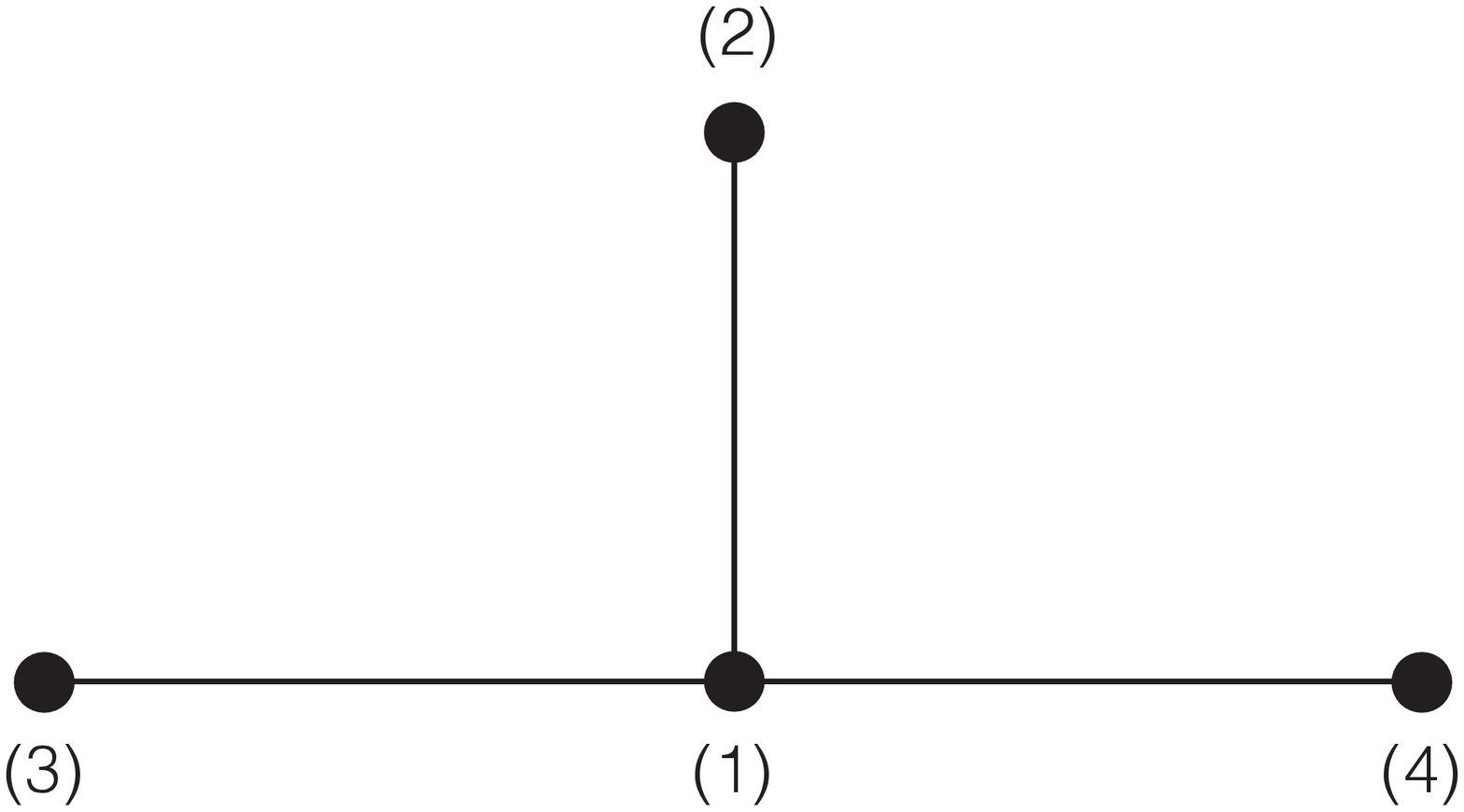,scale=0.3}	
	\label{fig:4_tree}
	\fcaption
	{
		A tree graph of length 4. This graph shows only physical output qubits of the $[[4,2,2]]$ graph code.
	}
}
\end{figure}

When $k>1$, one can find $n-k$ stabilizer generators of a graph code by iterating the above-mentioned process $k$ times.
In the $(j+1)$-th iteration, one can obtain $n-(j+1)$ operators those are commutable with the logical Z operator $\bar{Z}_{j+1}$ from the $n-j$ operators obtained in the $j$-th iteration.
For example, a $4$-node tree $T_4$ shown in Fig.~\ref{fig:4_tree} is a graph for the $[[4,2,2]]$ code~\cite{Cafaro:2014vm}\footnote{This is a normal graph that consists of only 4 physical output nodes unlike the graph shown in Ref.~\cite{Cafaro:2014vm}.}.
The adjacency matrix of this graph and the stabilizer generators of the graph state by this graph are respectively
\begin{equation}
\Gamma(T_4) = \left(
\begin{array}{cccc}
0 & 1 & 1 & 1 \\
1 & 0 & 0 & 0 \\
1 & 0 & 0 & 0 \\
1 & 0 & 0 & 0
\end{array}
\right)~\textrm{and}~
\mathcal{G} = \left(
\begin{array}{cccc}
X & Z & Z & Z \\
Z & X & I & I \\
Z & I & X & I \\
Z & I & I & X
\end{array}
\right).
\end{equation}
One can find there is a matrix $B$ of size $2\times 4$ such that $B\cdot \Gamma(T_4) = 0$, where each row vector is linearly independent,
\begin{equation}
B = \left(
\begin{array}{cccc}
0 & 1 & 1 & 0 \\
0 & 0 & 1 & 1 \\
\end{array}
\right),
\end{equation}
and the logical Z operators $\bar{Z}_i$ of the associated graph code, those are related to $B$, are $\bar{Z}_1 = I Z Z I$ and $\bar{Z}_2 = I I Z Z$.
First, one picks 3 commutable operators $\{XZZZ,~ZIIX,~IXXI\}$ commuting with $\bar{Z}_1$ by the above-mentioned procedure, and then for $\bar{Z}_2$, one obtains 2 commutable operators $XZZZ$ and $ZXXX$.
These last 2 operators are locally equivalent with the stabilizer generators of the $[[4,2,2]]$ code described as follows
\begin{equation}
\mathcal{G}^{[[4,2,2]]} = \left(
\begin{array}{cccc}
X & X & X & X \\ Z & Z & Z & Z
\end{array}
\right).
\end{equation}

A stabilizer group $\mathcal{S}_{\mathcal{C}}$ of a graph code $\mathcal{C}_G$ stabilizes a graph state $|G\rangle$, and therefore $\mathcal{S}_{\mathcal{C}}$ corresponds to a subgroup of the stabilizer group $\mathcal{S}_{|G\rangle}$ of $|G\rangle$, $\mathcal{S}_{\mathcal{C}}\subset \mathcal{S}_{|G\rangle}$.
Logical $X$ operators $\bar{X}$ of $\mathcal{C}_G$ also stabilize the graph state $\bar{X}(|0\rangle_L + |1\rangle_L) = (|1\rangle_L + |0\rangle_L)$, that is $\bar{X}\in\mathcal{S}_{|G\rangle}$.
By considering the fact that the dimensions of $\mathcal{S}_{\mathcal{C}}$ and $\mathcal{S}_{|G\rangle}$ respectively are $2^{n-k}$ and $2^{n}$, and the dimension of a group $G_{\bar{X}}$ generated by $\bar{X}$s is $2^k$, we can say that $\mathcal{S}_{|G\rangle}$ is equivalent with a product of $\mathcal{S}_{\mathcal{C}}$ and $G_{\bar{X}}$, 
\begin{equation}
\mathcal{S}_{|G\rangle} \equiv (\mathcal{S}_{\mathcal{C}}, G_{\bar{X}}).
\end{equation}

Note that after the first submission of this work, we recognized that the stabilizer group of a graph code is described in Refs.~\cite{Schlingemann:2001tv,Grassl:2002ca}. 
The resulting stabilizer groups by our method and their description are the same, but the principles of both are different as follows.
Our method makes use of the relation between stabilizers of a graph code/state and the logical Z operators of the graph code.
On the other hand, they apply the adjacency matrix of a normal graph to the images of the matrix $B$, $(k|\Gamma(G)\cdot k)$, where $B\cdot k = 0$.

\section{Conclusion}\label{sec:conclusion}

We have investigated a relation between a graph state and a graph code, both are defined by the same graph.
A graph state is a superposition of logical qubits of the related graph code, and a logical qubit of the graph code can be obtained by performing a projection to the graph state.
By using this relation, we have first argued that a special local Clifford operation acting on a graph state can be useful for searching locally equivalent stabilizer codes.
It is well known that a stabilizer code is transformed into a graph state (or a graph code) by local Clifford operations.
An application of the local Clifford operations transforms a graph state into another locally equivalent graph state, and the resulting graph state is projected to a logical qubit of the related graph code.
We will give details of this sketch in a forthcoming paper.

Second, we have provided a method to find stabilizer generators of a graph code.
To date, a graph code has not been described by a standard stabilizer formalism based on a Pauli group (or equivalently symplectic group), and therefore it has not been considered for a fault tolerant quantum computing.
Now, we believe that a graph code can make a greater role in a fault tolerant quantum computing than ever.

\nonumsection{Acknowledgements}
\noindent
This research was supported by Basic Science Research Program through the National Research Foundation of Korea (NRF) funded by the Ministry of Education (2014R1A6A3A01009674), the MSIP (Ministry of Science, ICT and Future Planning), Korea, under the ITRC (Information Technology Research Center) support program (IITP-2015-R0992-15-1017) supervised by the IITP (Institute for Information \& communication Technology Promotion) and the ICT R\&D program of MSIP/IITP (R0190-15-2030). 

\nonumsection{Reference}

\bibliography{reference}
\bibliographystyle{unsrt}

\end{document}